\documentclass[preprint]{aastex}
\usepackage{psfig}

\newcommand\mdot   {\hbox {${\dot M}$}}

\newcommand\pp     {$\pm$}
\newcommand\pers     {s$^{-1}$}
\newcommand\micros  {$\mu$s}

\newcommand\funit   {erg cm$^{-2}$ s$^{-1}$}

\righthead{The burst behavior of MXB 1659--298}
\slugcomment{Accepted for publication in ApJ: 18 October, 2001}

\begin{document}

\title{The burst behavior of the eclipsing low-mass X-ray binary MXB
1659--298}

\author{Rudy Wijnands\altaffilmark{1,4}, Michael
P. Muno\altaffilmark{1}, Jon M. Miller\altaffilmark{1}, Luc\'\i a
M. Franco\altaffilmark{2}, Tod Strohmayer\altaffilmark{3}, Duncan
Galloway\altaffilmark{1}, Deepto Chakrabarty\altaffilmark{1}}

\altaffiltext{1}{Center for Space Research, Massachusetts Institute of
Technology, 77 Massachusetts Avenue, Cambridge, MA 02139-4307, USA;
rudy@space.mit.edu}

\altaffiltext{2}{University of Chicago, 5640 S. Ellis Ave., Chicago IL 60637}

\altaffiltext{3}{Laboratory for High Energy Astrophysics, Goddard Space Flight
Center, Greenbelt, MD 20771}

\altaffiltext{4}{Chandra Fellow}

\begin{abstract}
We present a detailed study of the correlations between the burst
properties and the inferred mass accretion rate for the X-ray
transient MXB 1659--298. The bursts which exhibited oscillations were
observed when the source was at relatively high mass accretion rate,
similar to what has been seen for other sources. However, due to the
limited number of observations at lower mass accretion rates, no
bursts were observed at such accretion rates and it is still possible
that when MXB 1659--298 accretes at such low mass accretion rates,
bursts can occur which might still exhibit burst oscillations. No
clear correlations were found between the different burst properties
and accretion rate, in contrast to what has been found for KS
1731--260 and 4U 1728--34, but similar to what has been reported for
Aql X-1. However, this lack of correlation for MXB 1659--298 and Aql
X-1 might be due to the limited range of mass accretion rate observed
for those sources compared to KS 1731--260 and 4U 1728--34.
\end{abstract}


\section{Introduction \label{intro}}

Using the proportional counter array (PCA) on board the {\it Rossi
X-ray Timing Explorer} ({\it RXTE}), nearly coherent oscillations (or
burst oscillations) have now been found in nine\footnote{Evidence for
burst oscillations has also been reported for one burst of the
millisecond X-ray pulsar SAX J1808.4--3658 using the {\it BeppoSAX}
satellite; In 't Zand et al. 2001} low-mass X-ray binaries (LMXBs) out
of the more than fifty systems that exhibit type-I X-ray bursts (see,
e.g., Strohmayer 2001 for a review about burst oscillations). The
properties of these burst oscillations (coherence, strength,
amplitude, stability) suggest that they are most likely due to the
spin of the neutron star (e.g., Strohmayer, Zhang, \& Swank 1997a;
Strohmayer et al. 1996, 1998a, 1998b; Strohmayer \& Markwardt 1999;
Muno et al. 2000). The behavior of the burst oscillations is complex
and often their frequency increases during the decay of the burst
(e.g., Strohmayer et al. 1996; although a decreasing trend in certain
bursts have been reported by Strohmayer 1999, Miller 2000, and Muno et
al. 2000).  This increase is usually explained by assuming that at the
start of the burst the burning layer expands by several tens of
meters, causing the burning layer to slow down. Later on in the burst,
this layer relaxes back to the neutron star surface and spins up again
(see, e.g., Strohmayer et al. 1996). Recent theoretical calculations
pointed out that the theoretical expected values of spin-down
(assuming rigid rotation) are a factor of at least three smaller than
the frequency shifts observed (Cumming et al. 2001; see also Cumming
\& Bildsten 2000; see, however, Spitkovsky, Levin, \& Ushomirsky
2001). 

Clearly, the physical mechanism behind the burst oscillations and
their frequency evolution is still not well understood.  Also, it is
not clear how the behavior of the burst oscillations is influenced by
the other properties of the bursts (e.g., the burst profile, episodes
of radius expansion) or with the other properties of the source (i.e.,
the variations in the mass accretion rate (\mdot) onto the neutron
star surface). Recently, progress has been made in our understanding
of the correlations between the burst oscillation behavior and other
burst and source properties. However, the situation is complex and so
far a consistent picture, valid for all sources, has not yet emerged.

\subsection{Previous studies}

The accretion rate in neutron-star LMXBs is usually inferred from the
position of the sources in their X-ray color-color diagrams (CDs) or
hardness-intensity diagrams (HIDs). The low-luminosity sources trace
out an atoll-like shape (hence the name atoll sources), with the
different branches referred to as the banana branch (high inferred
accretion rate) and the island state (low inferred accretion rate; see
Hasinger \& van der Klis 1989 for the classification of the different
types of neutron star LMXBs). The changes in the burst properties as a
function of mass accretion rate has been studied before by several
others (e.g., van der Klis et al. 1990), however, we will only
summarize here the recent {\it RXTE} work about correlating the burst
oscillations with other burst properties and the source accretion
rate. We refer to Muno et al. (2000) for an elaborate description of
pre-{\it RXTE} work.  Muno et al. (2000) extensively studied the
behavior of the bursts and their oscillations with variations in the
inferred mass accretion rate (using a CD) for the X-ray transient KS
1731--260.  It was found that when this source was on its banana
branch (high inferred \mdot) that the X-ray bursts typically had short
decay times, were relatively bright, demonstrated clear episodes of
radius expansion (RE), and exhibited burst-oscillations. In contrast,
when the source was in its island state the bursts had longer decay
times, were weaker, showed no evidence for RE, and no burst
oscillations could be detected. Muno et al. (2000) identified the
banana branch bursts (with burst oscillations) as occurring in a
helium-rich environment and those in the island state (without
burst-oscillations) occurring in an environment which has a
considerable amount of hydrogen mixed in. The theoretical study of
Cumming \& Bildsten (2000) is consistent with these findings, in the
sense that large amplitude oscillations are more likely to be observed
in pure He-bursts than for mixed H/He-bursts.

Two other sources have been similarly studied: 4U 1728--34 (Franco
2001; van Straaten et al. 2001) and Aql X-1 (Fox et
al. 2001). However, different correlations were found between the
burst properties and variations in the mass accretion rate.  As with
KS 1731--260, the bursts observed for 4U 1728--34 did not exhibit
oscillations at low inferred \mdot, but only at larger \mdot~(Franco
2001; van Straaten et al. 2001). Moreover, not only are the burst
oscillations present only above a certain mass accretion rate, their
strength increases with \mdot~(Franco 2001). Unlike KS 1731--260, an
anti-correlation was found between the presence of RE episodes during
the bursts and \mdot, with not all bursts which exhibited oscillations
also experienced RE episodes. A final difference is that the brightest
bursts occurred on the banana branch for KS 1731--260 (Muno et
al. 2000) but in the island state for 4U 1728--34 (Franco 2001; van
Straaten et al. 2001).

Despite the fact that both sources behave differently with respect to
variations in mass accretion rate, for each source individually the
burst behavior seems to be correlated with \mdot. However, the results
obtained for Aql X-1 (Fox et al. 2001) demonstrates that such
correlations are not always observed. Only a limited amount of data
were obtained during the island state of this source and no X-ray
bursts were observed in this state; all bursts occurred at relatively
high mass accretion rate. The detection of burst oscillations during
one of these bursts is consistent with the findings that bursts
oscillations tend to occur preferentially at high mass accretion
rates. However, both bright and less bright (factor of $~$2 dimmer)
bursts occurred on roughly the same position in the CD, thus at
roughly the same mass accretion rate.  The same is true for bursts
with and without RE episodes.  The only burst exhibiting oscillations
has a clear RE episode, but a very similar burst which also exhibited
a RE episode showed no burst oscillations.  Clearly, currently the
burst behavior of Aql X-1 does not seem to be very well correlated
with the average mass accretion rate of the source just prior to the
onset of the bursts (Fox et al. 2001). However, we note that for Aql
X-1 only a limited amount of burst were observed and only at a limited
range of inferred mass accretion rate. Better and clearer correlations
might be observed when more bursts at a larger range of accretion
rates are observed for this source.  These three above discussed {\it
RXTE} studies, when combined with the previous studies done with data
obtained with other satellites (see, e.g., Lewin, van Paradijs, \&
Taam 1993 for an overview), demonstrate that presently no standard
picture which is valid for all sources exist regarding the correlation
between the burst behavior and the average mass accretion rate.

Very recently, it has been found (Muno et al. 2001) that the burst
oscillations observed for the systems with relatively high
burst-oscillation frequency ($\sim$500--600 Hz) are almost always
observed during RE bursts, whereas the burst oscillations for the
systems with low frequencies ($\sim$300 Hz) are approximately equally
found among bursts with and without RE episodes. Muno et al. (2001)
suggested that this might indicate that the burst properties change
differently as a function of \mdot~in the systems with high or low
frequencies.  To increase the number of sources (exhibiting burst
oscillations) for which the burst properties are correlated with the
mass accretion rate, we have undertaken a similar study for the
bursting LMXB MXB 1659--298.

\subsection{MXB 1659--298}

This system was first discovered by Lewin, Hoffman, \& Doty (1976).
Type-I X-ray bursts were reported from the system, demonstrating that
the compact object is a neutron star. The source exhibits periodic
X-ray dips and eclipses with a period of $\sim$7.1 hours, which can be
identified with the orbital period (Cominsky \& Wood 1984, 1989). In
April 1999, the source was detected after more then twenty years of
quiescence (in 't Zand et al. 1999). Observations of the source were
obtained with the proportional counter array (PCA) on board the {\it
RXTE} (e.g., Wachter, Smale, \& Bailyn 2000; Wijnands, Strohmayer, \&
Franco 2001), with {\it BeppoSAX} (Oosterbroek et al. 2001), and with
{\it XMM-Newton} (Sidoli et al. 2001).  With these observations, an
updated orbital ephemeris was obtained by Wachter et al. (2000) and
Oosterbroek et al. (2001).  Fourteen X-ray bursts were found in the
{\it RXTE} data and burst oscillations were discovered in eight of
them with a frequency of $\sim$567 Hz (Wijnands et al. 2001). Here we
report on the behavior of burst oscillations with respect to the other
burst properties (e.g., radius expansion bursts, rise time scales) and
the burst properties with respect to the variations in the mass
accretion rate onto the neutron star surface.

\section{Observations and analysis}

We used all currently available archival data for our analysis. Data
were obtained in the Standard 1 and 2 modes, and simultaneously using
an event mode (E\_125us\_64M\_0\_1s; 122 \micros~in 64 photon energy
channels covering the full {\it RXTE}/PCA energy range of 2--60
keV). The Standard 1 data (1/8 s time resolution and 1 energy channel)
were used to study the profiles of the bursts. The Standard 2 data (16
s resolution; 129 channels) were used to create the color-color
diagram (CD) and hardness-intensity diagram (HID) of the source. As
the soft color we used the count rate ratio between 4.1--7.5 keV and
2.9--4.1 keV, and as the hard color the ratio between 11.4--18.8 keV
and 7.5--11.4 keV. We have also made those diagrams using different
colors (although the limiting spectral resolution of {\it RXTE}/PCA
does not allow many variations of the colors), but the colors used
here gave diagrams which showed the different tracks the best.  The
event mode data were used to calculate 1/16--4096 Hz power spectra
which were used to investigate the rapid X-ray variability of the
source. The same data were also used to analyze the spectral evolution
during the bursts.

\section{Results}

\subsection{The long-term behavior of the source}

\begin{figure}[]
\begin{center}
\begin{tabular}{c}
\psfig{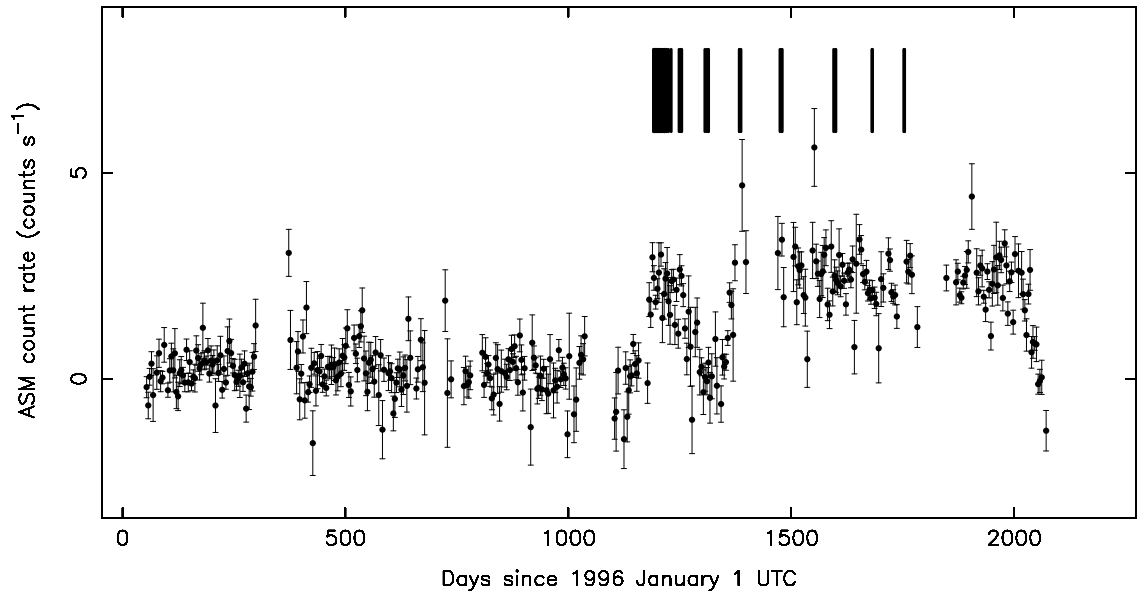}
\end{tabular}
\figcaption{The {\it RXTE}/ASM light curve (1--12 keV) of MXB
1659--298 up to 5 September 2001. The data points have been rebinned
into 4 day bins.  The solid lines are the times of the {\it RXTE}/PCA
observations.
\label{fig:asm} }
\end{center}
\end{figure}

We used the quick-look, one-day averaged {\it RXTE} All Sky Monitor
(ASM) data\footnote{The quick-look data can be obtained from
http://xte.mit.edu/ASM\_lc.html and is provided by the ASM/{\it RXTE}
team. See Levine et al. (1996) for a detailed description of the ASM.}
to determine the long-term X-ray light curve of the source
(Fig.~\ref{fig:asm}). The source was dormant for the first $\sim$1000
days of the {\it RXTE} mission but became active in April 1999 (in 't
Zand et al. 1999). The initial part of the outburst can be
characterized by a faster rise ($<$50 days) and slower decay time
scale ($>$100 days), however, instead of going back into quiescence,
the source exhibited a second outburst which continues until the end
of July 2001, after which the source started to decay again. It become
undetectable in the ASM around the end of August 2001. A pointed
observation with the {\it RXTE}/PCA on 7 September 2001, showed that
the source could still be detectable at a level of $\sim$5 mCrab
(2--60 keV), however, later observations on 14, 24, and 30 September
2001 showed that the source could not be detected with upper limits of
0.5--1 mCrab. In Figure~\ref{fig:asm}, we indicated when the {\it
RXTE}/PCA observations used in our analysis were performed.  As can be
seen, most of the data were obtained when the source was at relatively
high count rates, and only a few observations were performed when the
source was weaker.

\subsection{The variations in the mass accretion rate}

\begin{figure}[]
\begin{center}
\begin{tabular}{c}
\psfig{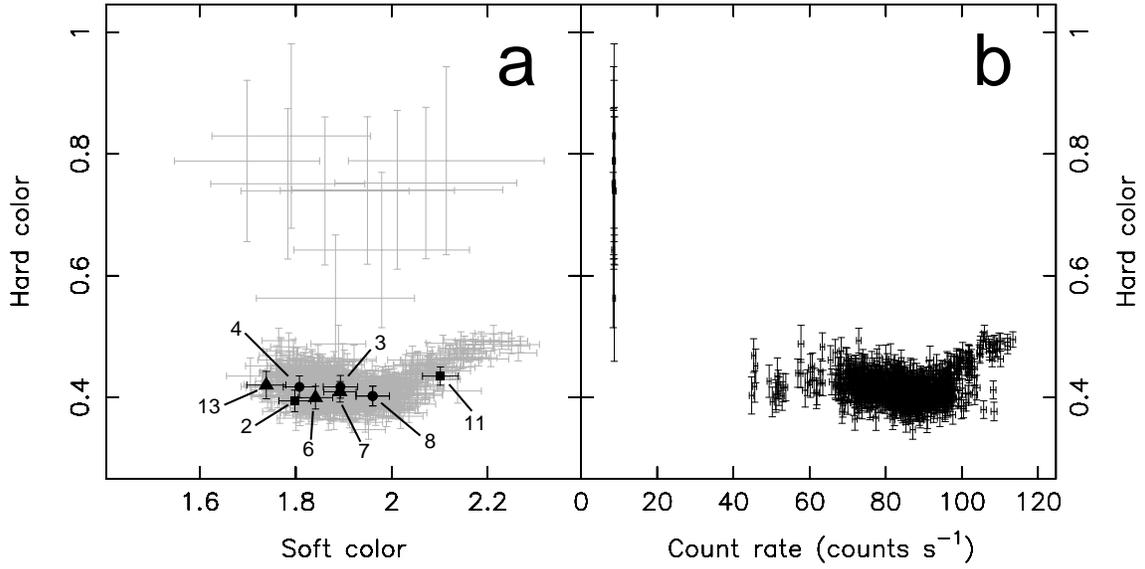}
\end{tabular}
\figcaption{The color-color diagram ({\it a}) and the
hardness-intensity diagram ({\it b}) of MXB 1659--298. The data were
from PCU 2 only with the dips and eclipses excluded. The time
resolution of the data is 256 seconds. In ({\it a}) the bursts are
indicated by number (according to Wijnands et al. 2001). Solid circles
indicate the bursts for which oscillations were detected in the tail
of the burst, the solid squares the ones for which oscillations were
found in the rise, and solid triangles those for which no oscillations
could be detected. The soft color is the count rate ratio between
4.1--7.5 keV and 2.9--4.1 keV, the hard color the ratio between
11.4--18.8 keV and 7.5--11.4 keV, and the count rate is between 2.9
and 18.8 keV.  The count rates were background subtracted but not
dead-time corrected.
\label{fig:cd_hid} }
\end{center}
\end{figure}

To correlate the burst behavior with the accretion rate, we have
created a CD and HID for MXB 1659--298.  However, because of the X-ray
dips and the eclipses, those diagrams are dominated by the very large
spectral variations during those events. Including those events the
diagrams were heavily contaminated and not useful to determine the
exact spectral state of the source.  Therefore, we decided to manually
remove all the data obtained during the dips or the eclipses (i.e., we
removed the data for which the count rate variations clearly showed
that the source was exhibiting dips or eclipses).  The resulting CD
and HID are shown in Figure~\ref{fig:cd_hid}. Clearly visible are two
branches in the diagrams, reminiscent of the atoll-shaped track of
other low-luminosity neutron star LMXBs; the lower branch can be
identified with the banana branch and the upper with the island
state. The rapid X-ray variability on the banana branch was weak
(several percent rms), as expected on this branch (Hasinger \& van der
Klis 1989). The variability could not be accurately determined in the
island state due to the very low count rate and limited amount of data
obtained in this state. However, the typical upper limits on the noise
properties during this state ($>$20\% rms) are fully consistent with
the expected timing properties for this state. This behavior clearly
demonstrate that MXB 1659--298 can be classified as an atoll source.

In Figure~\ref{fig:cd_hid}{\it a}, we indicated (by numbers; see
Wijnands et al. 2001 for the numbering of the bursts) when the bursts
occurred. To estimate the source colors during the time of the bursts,
we used 256-s of data just prior to them. We only show the bursts
which occurred outside the X-ray dips and the eclipses, because the
colors of the other bursts are strongly affected by the spectral
variations during those events. However, only very limited amount of
data were obtained during the island state and no bursts were observed
during the time the source was in this state; therefore, all bursts so
far observed in MXB 1659--298 occurred when the source was on the
banana branch, at relatively high inferred mass accretion rate (see
also Muno et al. 2001).

There is no clear correlation between the position of the source on
the atoll track at the time of the bursts and the presence (burst 2,
3, 4, 8, 11) or absence (burst 6, 7, 13) of burst oscillations
(Fig.~\ref{fig:cd_hid} {\it a}). Both bursts with and without burst
oscillations occurred at approximately similar locations in the
CD. Similarly, one of the two bursts (burst 2) with the burst
oscillations occurring only during the rising phase, occurred at
approximately the same positions in the CD as one of the bursts which
exhibited the burst oscillations during their tail (burst 4).

We also calculated the fluxes of the persistent emission (excluding
the dips and eclipses) just prior to the bursts. The spectra were
derived from the top Xenon layers of whichever detector (or
proportional counter unit [PCU]) was on during the
observations. Background subtraction and the creation of the response
matrices were performed using the standard techniques and FTOOLS
version 5.04. We fitted different functions to the spectral data, but
the obtained fluxes were very similar. We quote the obtained absorbed
fluxes in Table~\ref{tab:properties}. No correlation was found between
the burst properties and the 3--25 keV flux.

\subsection{Extra bursts}

\begin{figure}
\begin{center}
\begin{tabular}{c}
\psfig{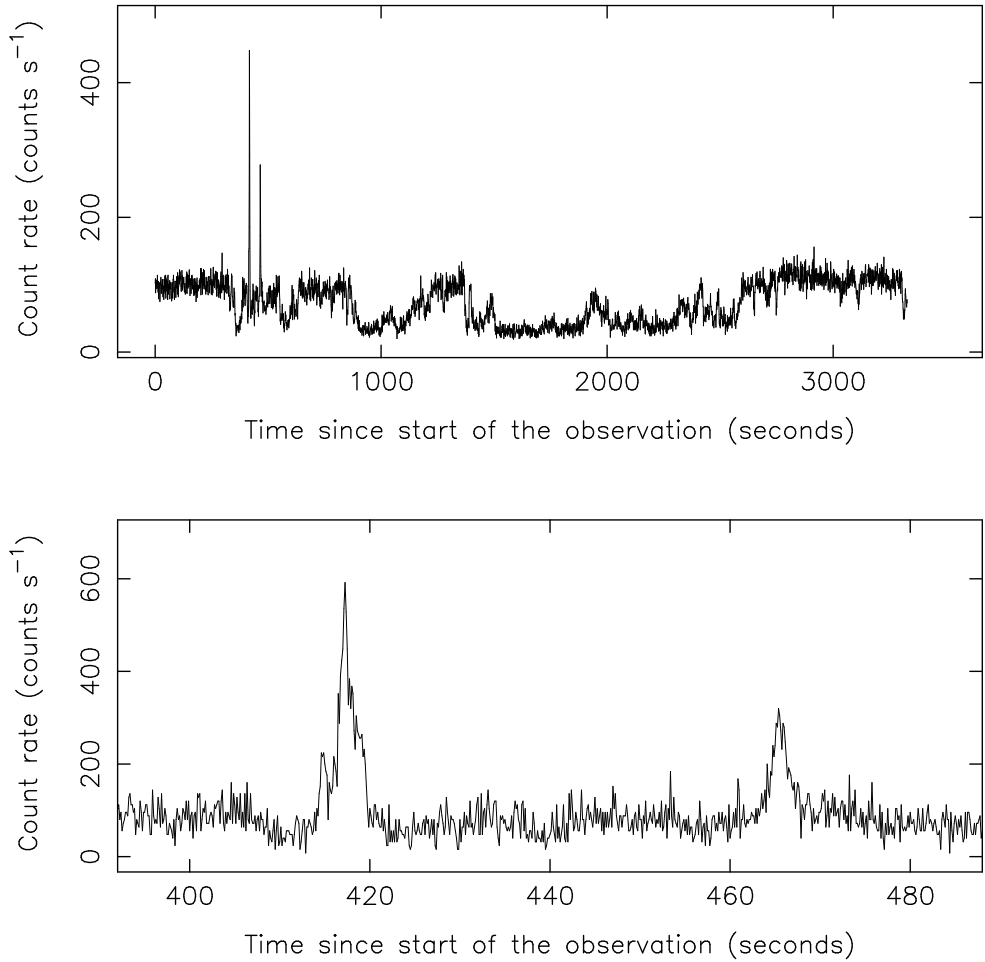}
\end{tabular}
\figcaption{The light curves of the bursts during observation
40050-04-09-01. The top panel shows the light curve of the total
observation, which clearly shows the dipping activity of the source;
the bottom panel shows a close-up of the two bursts. The count rates
are for 1 PCU, 2--60 keV, and are not background subtracted or dead
time corrected. The time resolution is 1 second in the top panel and
0.125 seconds in the bottom panel.
\label{fig:bursts_extra} } 
\end{center} 
\end{figure}

We report the detection of two extra bursts which were
missed\footnote{They were missed because of the large bin size (16 s)
used by Wijnands et al. (2001) to search for bursts, which is
considerably longer than the bursts duration. In the present study, we
used a bin size of 1/8 seconds which makes it unlikely that any more
bursts are missed.} by Wijnands et al. (2001;
Fig.~\ref{fig:bursts_extra}). No oscillations could be detected during
these bursts. Remarkably, they happened within only $\sim$50 seconds
of each other. Secondary bursts occurring $<$10 minutes after the
primary bursts have been reported before in several sources (see,
Lewin et al. 1993 for a discussion), which are thought to be due to
unstable burning of residual fuel which did not burn during the
primary bursts (see, e.g., Fujimoto et al. 1987). A similar
explanation might be valid for what we see for MXB 1659--298, although
a recurrence time of 50 seconds is very short. Sadly, detailed
investigations of these bursts (e.g., their spectra) are hampered by
the fact that they occurred during the X-ray dips
(Fig.~\ref{fig:bursts_extra}). Therefore, we do not discuss these
bursts further in our paper.

\subsection{Burst profiles}

\begin{figure}[]
\begin{center}
\begin{tabular}{c}
\psfig{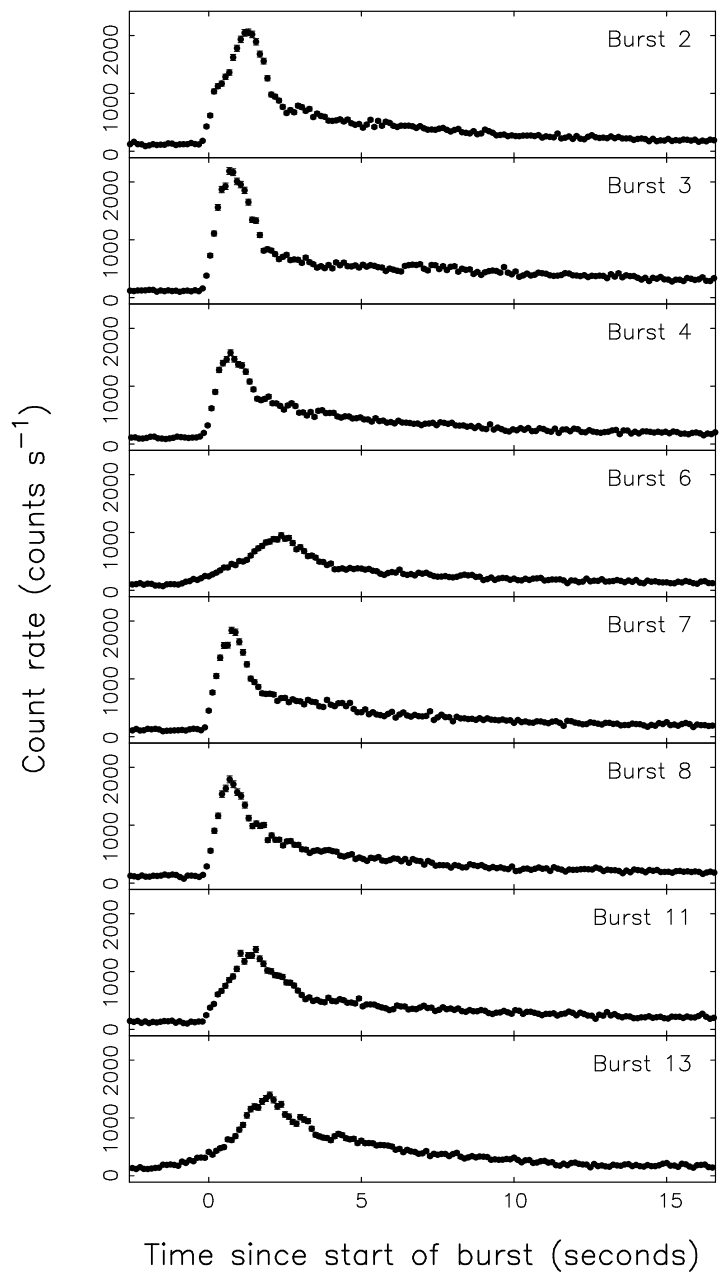}
\end{tabular}
\figcaption{The {\it RXTE}/PCA light curves of the bursts of MXB
1659--298. The data points are 1/8 seconds bins. The count rates are
for 2--60 keV, for 1 detector, and are not background subtracted or
dead time corrected. Time is from the start of the bursts, which is
defined as the time when the count rate reached a level of 25\% of the
peak count rate.  For bursts 2 and 11 the burst oscillations were
found only during the rise of the bursts. For burst 3, 4, and 8 the
oscillations were only found in the tail of the bursts. No
oscillations were found for bursts 6, 7, and 13.
\label{fig:burst_profiles} } 
\end{center} 
\end{figure}

The burst profiles of the bursts discussed in this paper (excluding
those that occurred when the source was dipping or eclipsed) are shown
in Figure~\ref{fig:burst_profiles}. From this figure it can be seen
that the bursts in MXB 1659--298 exhibit a wide variety of bursts
profiles. To quantify their behavior, we have determined the rise time
($t_{\rm rise}$, defined as the time it takes for the flux to increase
from 25\% to 90\% of the peak flux) and the decay time of the bursts
($t_{\rm decay, 1}$ and $t_{\rm decay, 2}$ , defined as the
exponential decay time; note that two successive exponential functions
were needed to fit the decay in the tail of the bursts correctly,
similar to what has been observed for, e.g., KS 1731--260 by Muno et
al. 2000, and Aql X-1 by Fox et al. 2001). The rise and decay times
are listed in Table~\ref{tab:properties}.  The bursts can be divided
in two groups; one with rise times shorter than 0.7 seconds, and the
other with rise times between 1 and 2 seconds. The two groups of
bursts in terms of their rise time seems to be correlated with the
burst oscillations occurring either in the tail or in the rise. Bursts
3, 4, and 8 have all a rise time of 0.5--0.6 second {\it and} exhibit
burst oscillations in the tail of the bursts; bursts 2 and 11 have
rise times of 1.0 and 1.4 seconds, respectively, and both show
oscillations during the rise of the bursts only.

Although this correlation seems to be strict, the two groups also
contain bursts which do not show oscillations. Burst 7 also has a
short rise time (0.7 seconds), but no oscillations have been detected
in this burst (i.e., not in the tail as might be expected).  Wijnands
et al. (2001) already pointed out that this burst has a unusual excess
in burst flux about 4 seconds after the burst peak, which is not
apparent for the other bursts which have short rise times
(Fig.~\ref{fig:burst_profiles}).  This burst flux excess occurs at
approximately the same time as when the burst oscillations are
expected to appear (as judged from the similar bursts 3, 4, and
8). The spectral properties of this burst (see
\S~\ref{section:spectra}) do not show any obvious change at the time
of the occurrence of this burst flux excess (Fig.
~\ref{fig:parameters_none} {\it top right}).

\begin{figure}[]
\begin{center}
\begin{tabular}{c}
\psfig{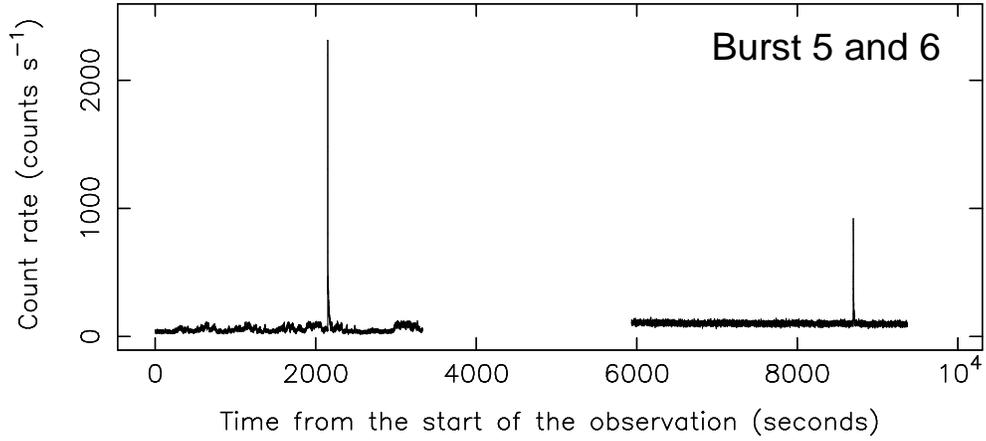}
\end{tabular}
\figcaption{The light curves of bursts 5 and 6. The count rates are
for 1 detector, 2--60 keV, and are not background subtracted or dead
time corrected. The time resolution is 0.25 seconds.
\label{fig:burst5_6} } 
\end{center} 
\end{figure}

In addition to burst 7, bursts 6 and 13 also do not exhibit
oscillations. It is intriguing that those bursts have the longest rise
times of all bursts in our sample (2.0 and 1.7 seconds,
respectively). Another remarkable fact about burst 6 is that it is the
weakest burst in our sample. This might be related to the fact that
this burst occurred only about two hours after the previous burst (see
Fig.~\ref{fig:burst5_6}; burst 5 occurred during dipping activity and
is not discussed in detail in this paper). However, this might also be
unrelated, because regular bursting of MXB 1659--298 with a interval
time of $\sim$2.5 hours had been observed before (Lewin et
al. 1976). Burst 13 is the longest burst in our sample: besides the
long rise time, the burst has a very long decay time ($t_{\rm decay,
1} \sim 3$, $t_{\rm decay, 2} \sim 15$). Besides the unusual profiles,
burst 6 and 13 are the only bursts in our sample which do not show
evidence for RE episodes (\S~\ref{section:spectra}). These unusual
properties of these two bursts might be related to the fact that they
do not exhibit burst oscillations, but the number of bursts is too
small to make definite conclusions and a chance coincidence cannot be
ruled out.

\subsection{The burst properties \label{section:spectra}}

\begin{figure}[]
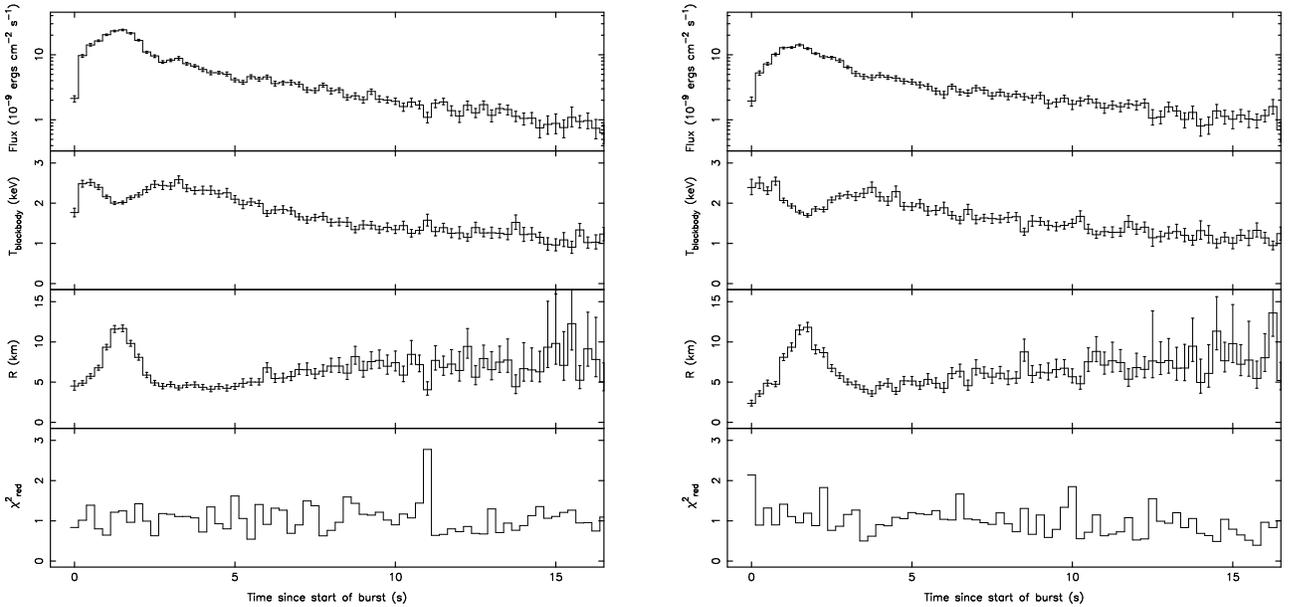

\begin{center}
\begin{tabular}{c}
\psfig{figure=f6a.eps,width=8cm}\hspace{1cm}\psfig{figure=f6b.eps,width=8cm}
\end{tabular}
\figcaption{The burst parameters versus time for the bursts which
exhibit burst oscillations only during the rise of the bursts (burst 2
[{\it left}] and burst 11 [{\it right}]). The error bars represent
90\% confidence intervals. The flux is the bolometric flux. The radii
are estimated for an assumed distance of 10 kpc.
\label{fig:parameters_rise} } 
\end{center} 
\end{figure}

\begin{figure}[]
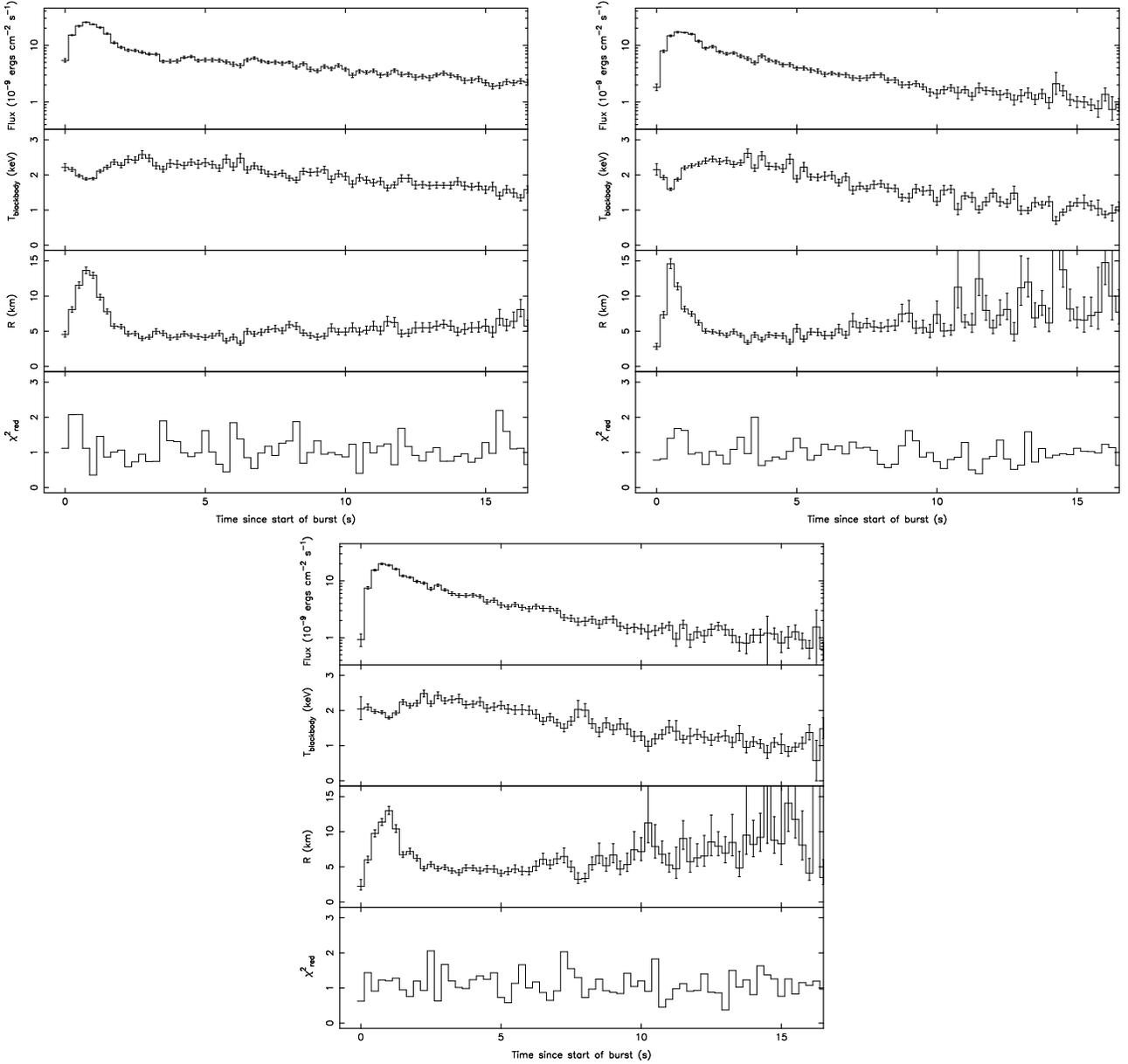

\begin{center}
\begin{tabular}{c}
\psfig{figure=f7a.eps,width=8cm}\hspace{1cm}\psfig{figure=f7b.eps,width=8cm} \\
\psfig{figure=f7c.eps,width=8cm}
\end{tabular}
\figcaption{The burst parameters versus time for the bursts which
exhibit burst oscillations only during the tail of the bursts (burst 3
[{\it top left}], burst 4 [{\it top right}], and burst 8 [{\it
bottom}]). The error bars represent 90\% confidence intervals. The
flux is the bolometric flux. The radii are estimated for an assumed
distance of 10 kpc.
\label{fig:parameters_tail} } 
\end{center} 
\end{figure}

\begin{figure}[]
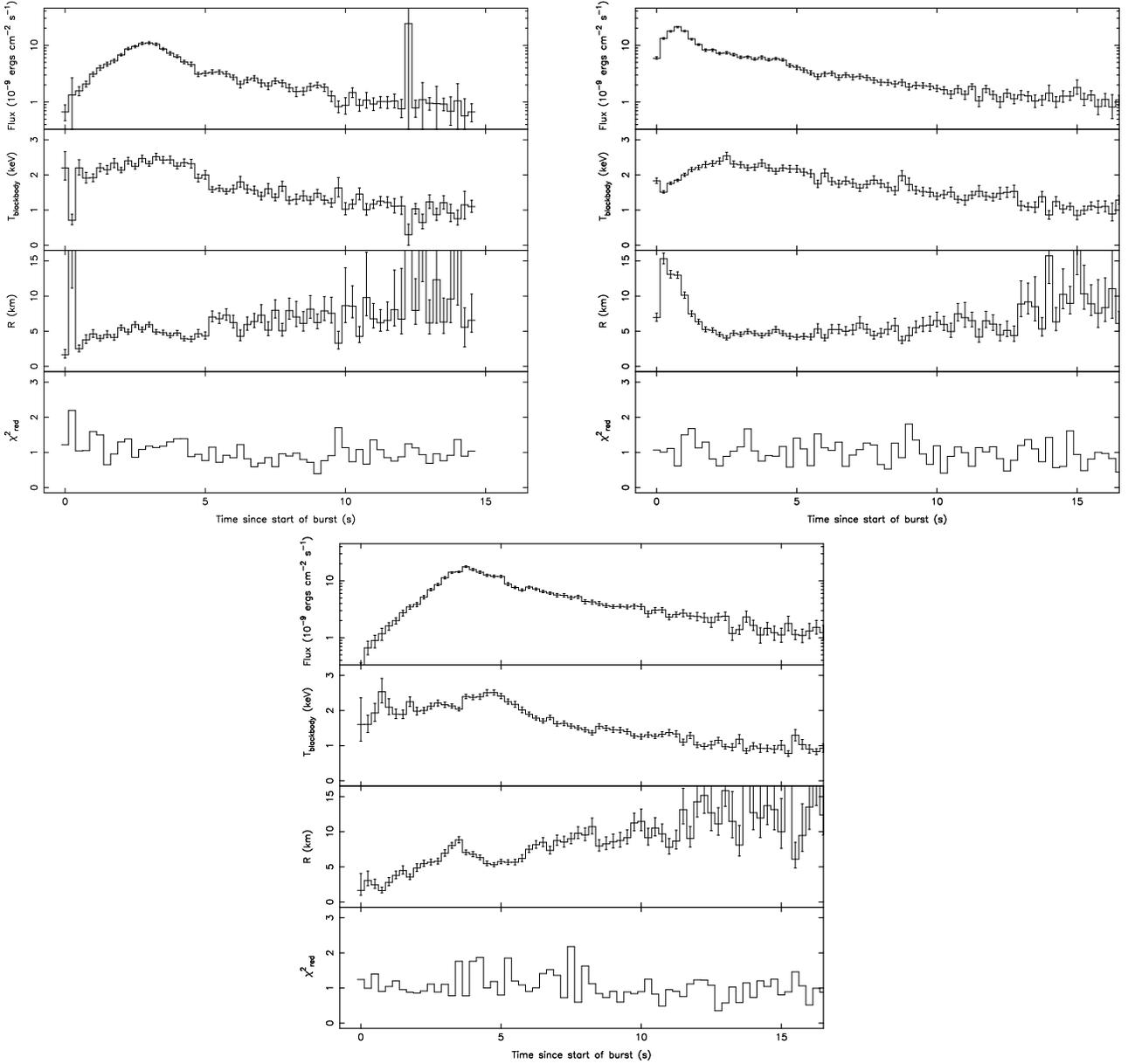

\begin{center}
\begin{tabular}{c}
\psfig{figure=f8a.eps,width=8cm}\hspace{1cm}\psfig{figure=f8b.eps,width=8cm} \\
\psfig{figure=f8c.eps,width=8cm}
\end{tabular}
\figcaption{The burst parameters versus time for the bursts which did
not exhibit burst oscillations (burst 6 [{\it top left}], burst 7
[{\it top right}], and burst 11 [{\it bottom}]). The error bars
represent 90\% confidence intervals. The flux is the bolometric
flux. The radii are estimated for an assumed distance of 10 kpc.
\label{fig:parameters_none} } 
\end{center} 
\end{figure}

We produced energy spectra for the bursts for each 0.25 second
interval from the event mode data. Once more, the bursts during the
dips and eclipses are left out of our analyses because their spectral
parameters are heavily contaminated by those events and no reliable
information about the intrinsic spectral behavior of those bursts
could be obtained. As background we used 15 seconds of data prior to
each burst. This assumes that the occurrence of the bursts did not
effect the overall properties of the persistent emission, which most
likely is an over-simplification of the situation. However, the
procedure is standard for analyzing the spectral evolution of bursts
and other, more complicated procedures give very similar results (see,
e.g., Muno et al. 2000 and Kuulkers et al. 2001 for a discussion). We
fitted the spectra between 2.5--20 keV with an absorbed blackbody (the
column density used was fixed to $0.7 \times 10^{22}$ cm$^{-2}$, which
was the mean value obtained from fits using a variable absorption).
From the model, the apparent blackbody temperature ($T_{\rm
blackbody}$) and the normalization (equal to the square of the
apparent radius $R$ of the emission region) are obtained. The
evolution of these parameters for each burst in our sample are shown
in Figures~\ref{fig:parameters_rise}--\ref{fig:parameters_none} (note
that the radius is normalized for a distance of 10 kpc; Muno et
al. 2001).  The resulting spectral parameters should be used with
caution, because it is unlikely that the emission during the bursts is
a pure blackbody (e.g., London, Taam, \& Howard 1984). We refer to
Muno et al. (2000) and references therein for a discussion about this.

From Figures~\ref{fig:parameters_rise} and~\ref{fig:parameters_tail}
it is clear that all bursts with burst oscillations exhibit a RE
episode during which the temperature decreases and the radius
increases. However, for the bursts which exhibit the oscillations in
the rise (Fig.~\ref{fig:parameters_rise}), this episode starts later
in the bursts ($>$1 seconds after the start of the burst) than for the
bursts which exhibit the oscillations in the burst tail
(Fig.~\ref{fig:parameters_rise}; $<$1 seconds). Although we have only
a limited number of bursts this correlation is strict. In the bursts
which exhibited burst oscillations in the tail, the oscillations only
appeared after the radius expansion episode already had ended.

Of the three bursts which do not exhibit burst oscillations
(Fig.~\ref{fig:parameters_none}), only one (burst 7) exhibits a RE
episode. However, this RE episode begins very early on in the burst
and this unusual behavior might be related to its lack of burst
oscillations. However, as already explained above, this burst also
exhibits a slight count rate excess approximately 4 seconds after the
start of the burst (see Fig.~\ref{fig:burst5_6}; see also Wijnands et
al. 2001). It is of course possible that all unique properties of this
burst are caused by the same underlying mechanism.

Burst 4 was identified by Wijnands et al. (2001) as unusual because of
the frequency behavior of the oscillations during this burst. Normally
when the oscillations were found in the tail of the bursts, their
frequency increased slightly by 0.5--1 Hz (over a time span of a few
seconds). For burst 4, the oscillations also increased first by
$\sim$1 Hz before they died away. However, about 4 seconds later they
reappeared again but at a frequency almost 5 Hz larger (see Fig. 3 of
Wijnands et al. 2001). Our spectral analysis of this burst
(Fig.~\ref{fig:parameters_tail} {\it top right}) does not show
anything unusual at the time of the reappearance of the
oscillations. The reason for this reappearance of the oscillations in
this particular burst and not in the other bursts remains therefore
elusive.

\section{Discussion}

We have made a detailed study of the burst behavior of the neutron
star X-ray transient MXB 1659--298.  Due to the limited amount of
data, bursts were only observed at relatively high inferred mass
accretion rates. Thus, a detailed investigation of the burst behavior
versus accretion rate as had been done for KS 1731--260 (Muno et
al. 2000) and 4U 1728--34 (Franco 2001; van Straaten et al. 2001)
could not be performed. However, consistent with those studies, we
find that at high inferred mass accretion rates (on the atoll source
banana branch), the bursts can exhibit burst oscillations, although
several bursts were observed at similar accretion rates, which do not
exhibit such oscillations.

Muno et al. (2001) briefly checked the color-color diagrams of all the
burst oscillation sources known (including MXB 1659--298, but
excluding MXB 1743--29 because source confusion made it very difficult
to isolate the persistent emission from this source), and they
reported that the bursts which exhibited oscillations all occurred at
relatively high mass accretion rates. However, due to the lack of
observations at low accretion rate for all sources besides KS
1731--260 and 4U 1728--34 (Muno et al. 2000; Franco et al. 2001; van
Straaten et al. 2001), it is possible that burst occurring at
relatively low mass accretion rates in those sources, might still
exhibit such oscillations. In this respect it is interesting to note
that for the X-ray transient SAX J1808.4--3658, burst oscillations
might have been seen (in 't Zand et al. 2001), but only when the
source was at very low luminosity (too low to be detected with the
Wide Field Cameras aboard {\it BeppoSAX}).  Therefore, if these
oscillations are due to the same phenomenon as the burst oscillations
seen in the other sources, then for SAX J1808.4--3658 the burst
oscillations are seen when the source is in its low-luminosity state,
contrary to what has been observed so far for the other
sources. However, it has been suggested (Psaltis 2001) that the
oscillations seen for SAX J1808.4--3658 are different from the burst
oscillation phenomenon and that they might be related to same
mechanism which produces the millisecond pulsations in the persistent
X-ray emission of this source (Wijnands \& van der Klis 1998).

Franco (2001) noticed that for 4U 1728--34, the bursts with
oscillations in the decay phase only occurred at mass accretion rates
lower than those bursts for which the oscillations were seen either
throughout the bursts or only during their rising phase. For MXB
1659--298, we do not see such a correlation: bursts with the
oscillations in the decay phase (e.g., bursts 3 and 4) can occur at
the same mass accretion rates as those with the oscillations only in
the rise (e.g., burst 2). It is interesting to note that the burst
which occurred at the highest inferred mass accretion rate (burst 11),
exhibited oscillations only during its rising phase. Currently, not
enough data are available for any of the burst oscillation sources to
determine whether the correlation found by Franco (2001) was a
coincidence; that it only applies to 4U 1728--34; or that it is valid
for more burst oscillation sources.
 
Besides the burst oscillations, the correlations of the other burst
properties (e.g., the rise and decay times, radius expansion episodes)
with the mass accretion rate for MXB 1659--298 are also not so clean
as for KS 1731--260 (Muno et al. 2000) and 4U 1728--34 (Franco 2001;
van Straaten et al. 2001).  For example, for MXB 1659--298, bursts
with and without episodes of radius expansion can occur at very
similar mass accretion rate. In this respect, MXB 1659--298 is more
similar to Aql X-1 (Fox et al. 2001) than to KS 1731--260 or 4U
1728--34. However, the bursts in both Aql X-1 and MXB 1659--298 have
so far only been detected at a rather limited mass accretion rate
range (especially compared to the other two sources). It might be
possible that in such a small range, the burst properties are not so
well correlated with the mass accretion rate, but that cleaner
correlations will arise when bursts are observed over a larger range
of mass accretion rates.

Muno et al. (2000) divided the bursts in KS 1731--260 in two distinct
groups: those bursts which exhibited burst oscillations, high peak
fluxes, short decay times, and episodes of radius expansion, and those
bursts which did not show oscillations, had low peak fluxes, had
longer decay times, and exhibited no episodes of radius
expansion. Muno et al. (2000) associated the first group with helium
rich bursts and the second group with hydrogen rich bursts.  The MXB
1659--298 bursts are not completely consistent with this division,
although they follow a similar trend for those which exhibited bursts
oscillations: in general those bursts have high peak fluxes (except
for burst 11 which is relatively dim), have episodes of radius
expansion, and have relatively short decay times. The bursts which do
not exhibit burst oscillations do not follow the KS 1731--260 division
so strictly. Burst 6 is the dimmest in our sample and has no radius
expansion episode, but it does have a relatively short decay time;
burst 7 has no oscillations but it is rather bright and has a short
decay time; burst 13 has no RE episode and it has the longest decay
time of all our bursts, but it is still as bright as the bursts which
exhibited burst oscillations.

Despite these differences between the burst behavior in KS 1731--260
and MXB 1659--298, those two sources are more similar to each other
than MXB 1659--298 is to 4U 1728--34.  In 4U 1728--34, the bursts
which exhibited burst oscillations are equally like to have or not to
have RE episodes (Franco 2001; van Straaten et al. 2001). All the
bursts in our example which show oscillations also exhibited RE
episodes. The burst oscillations in 4U 1728--34 occur in the dimmest
bursts, again contrary to what we observe for MXB 1659--298.  The fact
that MXB 1659--298 is more similar to KS 1731--260 in its bursting
behavior than to 4U 1728--34 might be related to the fact that the
burst oscillation frequency of KS 1731--260 and MXB 1659--298 are very
similar (524 Hz vs. 567 Hz), compared to 363 Hz for 4U 1728--34.

Recently, Muno et al. (2001) noticed that for the ``fast oscillators''
(those sources which have burst oscillations with frequencies above
500 Hz), the burst oscillations are predominantly observed in bursts
which also exhibit radius expansion episodes. For the ``slow
oscillators'' (burst oscillations frequency $\sim$300 Hz), the
oscillations are equally likely to be found in bursts with and without
radius expansion episodes. The bursts reported for MXB 1659--298 in
the present study were also used by Muno et al. (2001), so we cannot
conclude anything further from our present work about this
correlation. However, Muno et al. (2001) included all the bursts which
occurred during the dips and the eclipses, so their numbers for MXB
1659--298 are possibly effected by the effects of those events on the
measured bursts parameters (e.g., if the dips prevent one from seeing
oscillations or if they change the apparent spectral parameters of the
bursts). If we only use the bursts outside the dips and eclipses, the
correlations found for this source become even stronger. All five
bursts which exhibit burst oscillations also have a radial expansion
episode. Two of the three bursts which do not show oscillations do not
exhibit radial expansion, and only one burst remains which exhibits
radius expansion but no burst oscillations.

When comparing the accretion history of the fast rotators with the
slow rotators (as defined by Muno et al. 2001), a remarkable, although
not a strict, correlation can be found. All of the three known slow
rotators (4U 1728--34, 4U 1702--429, 4U 1916--053) are persistent
sources. In contrary, at least four out of the six fast rotators are
transients (4U 1608--52, MXB 1659--298, Aql X-1, and KS 1731--260) and
only one source is clearly persistent (4U 1636--53; the other source
still has conclusively to be identified with MXB 1743--298, which
might be a persistent source). This difference between the fast and
slow rotators might be related to their different burst behavior as
observed by Muno et al. (2001).  The neutron star crust and interior
will be different for the transient systems compared to the persistent
systems. The neutron stars in the transients are most likely cooler
than in the persistent systems, which will affect their burst behavior
(e.g., Fujimoto et al. 1984; see Lewin et al. 1993 for a discussion
and further references).

\acknowledgments

This work was supported by NASA through Chandra Postdoctoral
Fellowship grant number PF9-10010 awarded by CXC, which is operated by
SAO for NASA under contract NAS8-39073.  This research has made use of
data obtained through the HEASARC Online Service, provided by the
NASA/GSFC.

\clearpage

\clearpage

\begin{deluxetable}{ccccccccc}
\rotate
\tablecolumns{9}
\tablewidth{0pt}
\tablecaption{The properties of the X-ray bursts
\label{tab:properties}}
\tablehead{
\#$^a$    & Peak count rate$^a$       & Peak flux$^b$     & $t_{\rm rise}^c$ & $t_{\rm decay, 1}^d$ & $t_{\rm decay, 2}^d$ & Osc.$^a$ & RE episode & Persistent flux$^e$\\
          & (counts \pers~PCU$^{-1}$) & (10$^{-9}$ \funit)& (s)              & (s)                  & (s)                  &          &            & (10$^{-10}$ \funit)}
\startdata
2         & 2060                      & 24.2\pp0.8        & 1.0              &  0.51\pp0.04         &  6.9\pp0.3           & Rise     & Yes        & 8.9\\
3         & 2190                      & 25.5\pp0.5        & 0.6              &  0.57\pp0.03         & 12.5\pp0.3           & Tail     & Yes        & 8.9\\
4         & 1580                      & 17.1\pp0.6        & 0.5              &  0.50\pp0.04         &  7.3\pp0.2           & Tail     & Yes        & 8.6\\
6         &  950                      & 10.9\pp0.6        & 2.0              &  0.7\pp0.2           &  5.9\pp0.4           & No       & No         & 8.1\\
7         & 1840                      & 21.1\pp0.7        & 0.7              &  0.37\pp0.02         &  6.8\pp0.2           & No       & Yes        & 9.3\\
8         & 1800                      & 20.0\pp0.7        & 0.5              &  1.0\pp0.2           &  7.8\pp0.6           & Tail     & Yes        & 9.7\\
11        & 1385                      & 14.2\pp0.5        & 1.4              &  0.98\pp0.07         &  9.4\pp0.4           & Rise     & Yes        & 10.6\\
13        & 1520                      & 17.7\pp0.7        & 1.7              &  3.0\pp0.2           & 15\pp2               & No       & No         & 5.8\\
\enddata

\tablenotetext{a}{See Wijnands et al. 2001 for the numbering of the bursts, their peak count rate (2--60 keV), and if and
when the burst ocillations occured.}

\tablenotetext{b}{The bolometric peak flux of the bursts.}

\tablenotetext{c}{The rise time of the bursts. Defined as the time it
takes for the flux to increase from 25\% to 90\% of the peak flux.}

\tablenotetext{d}{Exponential decay times of the bursts; $t_{\rm
decay, 1}$ is the decay time for the first exponential and $t_{\rm
decay, 2}$ for the second one. }

\tablenotetext{e}{Absorbed 3--25 keV flux, just prior to the onset of
the bursts. The model used to obtain those fluxes was an absorbed
cut-off power-law with photon index of 1--1.5 and a cut-off energy of
5--6 keV. The assumed column density was $0.2 \times 10^{22}$
cm$^{-2}$. }

\end{deluxetable}

\end{document}